\RequirePackage{fix-cm}
\documentclass[natbib]{svjour3}                   
\smartqed  
\usepackage{graphicx}

\begin{document}

\title{Quantitative characterization of pore structure of several biochars 
with 3D imaging
}

\titlerunning{Characterization of biochars with 3D imaging}        

\author{Jari Hyv\"aluoma         \and
        Sampo Kulju \and
        Markus Hannula \and
        Hanne Wikberg \and
        Anssi K\"alli \and
        Kimmo Rasa
}


\institute{Jari Hyv\"aluoma \at
              Natural Resources Institute Finland (Luke), FI-31600 Jokioinen, Finland \\
              \email{jari.hyvaluoma@luke.fi}           
           \and
           Sampo Kulju \at
              Natural Resources Institute Finland (Luke), FI-31600 Jokioinen, Finland
              \and
           Markus Hannula \at
             BioMediTech Institute and Faculty of Biomedical Sciences 
             and Engineering,                   
             Tampere University of Technology, 
             Tampere, Finland
           \and
           Hanne Wikberg \at
              VTT Technical Research Centre of Finland Ltd., 
              P.O.Box 1000, FI-02044 VTT, Finland 
              \and 
           Anssi K\"alli \at
              VTT Technical Research Centre of Finland Ltd., 
              P.O.Box 1000, FI-02044 VTT, Finland 
              \and
           Kimmo Rasa \at
              Natural Resources Institute Finland (Luke), FI-31600 Jokioinen, Finland
}

\date{Received: date / Accepted: date}

\maketitle

\begin{abstract}

Pore space characteristics of biochars may vary depending on the used 
raw material and processing technology. Pore structure has significant 
effects on the water retention properties of biochar amended soils. 
In this work, several biochars were characterized with three-dimensional 
imaging and image analysis. X-ray computed microtomography was used to 
image biochars at resolution of
1.14 $\mu$m and the obtained images were analysed for porosity, pore-size
distribution, specific surface area and structural anisotropy. In
addition, random walk simulations were used to relate structural anisotropy
to diffusive transport. Image analysis showed that considerable part of the
biochar volume consist of pores in size range relevant to hydrological
processes and storage of plant available water. Porosity and pore-size
distribution were found to depend on the biochar type and the structural 
anisotopy analysis showed that used raw material considerably affects the 
pore characteristics at micrometre scale. Therefore attention should be paid 
to raw material selection and quality in applications requiring optimized 
pore structure.

\end{abstract}

\section{Introduction}
\label{intro}

In recent years biochar application to soils has been widely promoted due 
to its positive influences on the soil structure and fertility, as well as the 
possibility to sequestrate carbon into soils. Biochar additions change both
physical and chemical characteristics of soil and can thus affect
soil functions in many ways. One of the most commonly claimed benefit of 
biochar use as soil amendment is improved water retention properties. While 
main part of the 
studies exploring biochar amendments have ensued in positive results, 
there are also some reports on minor or even negative impacts on the water 
retention properties and other soil functions (see, e.g., 
\citet{Mukherjee14,Jeffery15,Novak16} and references therein). 

Knowledge of the underlying mechanisms how biochars affect soil water retention
is still incomplete \citep{Jeffery15}. Biochar may influence water retention
in direct or indirect way \citep{Verheijen10}. According to the direct mechanism
water is stored and 
held in the biochar pores. Biochars are typically multiscale porous materials 
containing a wide range of pore sizes. Direct
mechanism has been questioned on basis that majority of porosity would be
in the nanopore regime and thus the water stored in biochars not plant 
available
\citep{Verheijen10}. Recent studies suggest that while majority of 
biochar surface area is provided by nanopores, they only have a minor impact 
on the total porosity \citep{Brewer14,Gray14}.

The indirect mechanism suggest that interaction between soil and biochar 
alters the size, shape and arrangement of macro-aggregate pore system 
\citep{Yu16}, which affect soil hydraulic properties. Indirect evidences, 
like changes in aggregate stability, address that biochar may contribute 
significantly on soil structural development, although mechanisms behind these 
interactions are still unclear. 

It is likely that the actual influence of biochar additions on soil water 
retention is a combination of direct and indirect mechanisms, the latter 
being highly dependent on soil properties. In order to understand
the mechanisms involved, there is a clear need for better 
characterization of biochar porosity at micrometre scale \citep{Kinney12}.  

Structural properties of biochars can vary significantly depending on 
feedstock, processing method and process conditions. For example, a 
number of lignocellulosic feedstock (e.g., wood-based materials, straw 
and other agricultural by-products), crop processing by-products, animal 
manure, and municipal or industrial organic wastes have been subjected to 
pyrolysis or hydrothermal carbonization (HTC) \citep{Kambo15} to produce 
biochars or hydrochars
(for compactness hereinafter all chars are referred to as `biochars').  
Beside the raw material selection and process type, various process parameters,
most notably temperature, affect the biochar quality. In general, increase in
processing temperature generates biochars with higher total porosity
\citep{Brewer14}.

X-ray tomography is a non-destructive three-dimensional (3D) imaging method, 
which
together with image analysis provide means for characterization
and quantification of porous media. Modern microtomography scanners enable
imaging with micron and even sub-micron resolution whereby porous
materials can be studied at scales relevant for soil hydraulic 
properties \citep{Jeffery15}. In contrary, nanoscale
pores important for chemical sorption are not visible in tomographic
images. However, even in such applications larger pores are necessary as they
provide rapid connections from external surfaces to micropores.

Studies using tomographic techniques in biochar characterization are still 
fairly rare. \citet{Bird08} studied series of pyrolysed pine wood biochars at
resolution of 21 $\mu$m and analysed images for porosity, pore-size 
distribution and pore connectivity. \citet{Jones15} imaged 
cotton hull biochars at 4 $\mu$m resolution
and determined porosity as a function of pyrolysis temperature.
\citet{Conte15} imaged poplar biochar at 0.74 $\mu$m resolution and 
\citet{Jeffery15} hay biochar at 2.56 $\mu$m resolution. 
\citet{Schnee16} imaged wood and Miscanthus biochars at 5.67 $\mu$m 
resolution and studied their potential as microbial habitat.
\citet{Quin14} studied oil mallee biochar and imaged soils amended 
by this biochar. 

There are several alternative techniques which have been used in 
characterization of porous media (in the context of biochars, see, e.g., 
\citet{Baltrenas15,Rawal16}). 
Scanning electron microscopy provides high-resolution information
about the surface morphology but does not enable quantitative analysis
of the 3D pore structure.
Gas adsorption methods are widely used to determine surface area and pore-size 
distribution of biochars but they are not suitable for larger pore sizes 
relevant for soil amendments. Mercury intrusion porosimetry on the other hand 
can be
used to determine pore-size distribution at micrometre scale. Feasibility of
this method is decreased by the underlying assumptions which are used in
interpretation of the results \citep{Giesche06}. 
These assumptions are not valid for many 
realistic materials. NMR cryoporosimetry determines the pore-size distribution 
by observing the depressed melting point of a confined frozen liquid. This 
method is suitable for pore diameters upto ca. 1 $\mu$m, which excludes main 
part of the porosity relevant for water retention properties. Compared to these
techniques, x-ray microtomography enables versatile characterization of
pore space via image analysis and provides information about pores in size range
important for soil amendment purposes.

In this article the pore characteristics of several biochars are studied.
X-ray microtomography is used to reveal the internal structure of biochars
and the pore space is quantified by image analysis. Accurate imaging
at 1.14 $\mu$m resolution provides information about pore sizes relevant to 
the water retention properties of biochars. Our objective is to characterize
several biochars to find out how much variation there can be in the internal
structure of different biochars. Biochars produced from realistic (i.e.
heterogeneous) raw materials are used in this study, which allows observation 
of structural variation caused by material heterogeneity. Biochars produced
by slow pyrolysis and HTC are included in this study. While our work is 
motivated by use of
biochars as soil amendments in order to improve the water retention properties 
of soils, the results can be useful for other applications relying on transport,
storage and sorption properties of biochars.

\section{Materials and methods}

\subsection{Imaging}

Zeiss Xradia MicroXCT-400 (Zeiss, Pleasanton, USA) device was used in the 
x-ray computed microtomography imaging. All samples were imaged similarly. 
A 20$\times$ objective was used with binning 2 that resulted in pixel size 
of 1.14 $\mu$m. Projection angle was 360$^\circ$ with 1600 projections. 
Source voltage was 40 kV, source current 250 $\mu$A, and exposure time 3 
seconds. Filters were not used in the imaging. The reconstruction was done 
from the projections with Zeiss XMReconstructor software, which utilizes the 
filtered back projection algorithm. A kernel size of 0.7 was used in the 
smoothing filter. 

\subsection{Image processing}

X-ray tomographic reconstruction gives a 3D grid representation of the 
x-ray attenuation coefficient of the imaged sample as a grey-scale image. 
In order to analyse the pore space of the sample, images were converted 
into a binary representation where sample is segmented into two classes,
i.e., pore and solid voxels. This processing was done in several
steps.

First a region of interest (ROI) was selected such that maximal volume of the
sample was taken into the subsequent steps. The imaged biochar grains were 
small and in some cases single scan contained several distinct grains. 
Arbitrary shaped ROIs were allowed and they could also consist of several 
non-connected parts. Next the grey-scale image was filtered with 
a 3D median filter with radius 2.

Actual image segmentation was performed using a modified version of Otsu's 
automatic global thresholding algorithm \citep{Otsu79}. 
Otsu's algorithm is based on assumption of bimodal grey-scale histogram 
so that image consists of two distinct components (object and background; 
here pores and solids). Segmentation threshold is selected to minimize the 
weighted sum of variances,
\begin{equation}
\sigma_w^2 = w_p(t)\sigma_p^2(t) + w_s(t)\sigma_s^2(t).
\end{equation}
Above $w_i(t)$ are weights and $\sigma_i^2(t)$ variances of the two classes
($p$ = pores and $s$ = solids) for a given threshold value $t$.

Segmentation results obtained by Otsu's algorithm 
turned out to be unsatisfactory as the pore volume was exaggerated in 
the segmented image. Recently \citet{Hapca13} introduced a new segmentation 
method which adds a pre-classification step to the standard Otsu's algorithm.
In their method
voxels certainly belonging to solid class are first selected and Otsu's 
algorithm is then applied to rest of the voxels. In practice Hapca and 
co-workers used original Otsu's threshold as the preclassification threshold. 
This method was developed for soils which consist of several solid components 
with significantly varying density. The situation is quite different for 
biochar samples with much more homogeneous solid material density. Therefore 
it is not surprise that this method, in turn, underestimated the pore volume. 
We decided 
to use the average of the threshold values given by the standard Otsu method 
and the modified method by Hapca et al. in segmentation of our biochar images. 
The use of average threshold value led to segmentation which corresponded 
very well with manual segmentation using visually selected threshold 
value. While there is no
theoretical basis for using this method, we nevertheless settled on the 
average threshold approach instead of manually selected threshold to ensure 
objectivity of the segmentation process for different samples thus avoiding
operator-dependent bias in the results.

After segmentation, binary images were filtered with a majority filter  
(radius 2). Finally isolated solid objects with volume less than 1000 voxels 
were removed from the images. The resulting binary image was used in the actual 
image analyses with one exception; in anisotropy analyses (described below) 
the denoised grey-scale image was used.

\subsection{Image analysis}

Image analysis was performed to determine porosity, specific surface area, pore 
size distribution and degree of anisotropy of the imaged samples.

Porosity is the fraction of pore volume in the imaged sample and is simply 
obtained as quotient between the number of pore voxels and the total number of 
voxels in the ROI.

Specific surface area was calculated using a method based on Minkowski 
functionals \citep{Vogel10}. Surface area is related to the second Minkowski 
functional and can be determined from the number of pore-to-solid transitions
in the image. These transitions were counted within a $2 \times 2 \times 2$ 
basic cube by evaluating voxel configuration in such cube at every location 
in the sample. Specific surface area is obtained by determining the frequency
of the 256 possible configurations and using classical stereological formula. 
Detailed description of the algorithm is given by \citet{Vogel10}.

Pore-size distributions were calculated using a method based on mathematical 
morphology \citep{Horgan98}. In mathematical morphology image is probed with a 
structuring element which is a small test set. Here, a sphere was used 
as the structuring element. The two basic morphological 
operations are erosion and dilation. Erosion of the set $X$ (in the present 
case all pore voxels) by the structuring element $S$ is
\begin{equation}
\mathcal{E}_S(X) = \left\{ \mathbf{r} : S_\mathbf{r} \subset X  \right\},
\end{equation}
i.e., all centres $\mathbf{r}$ of the structuring elements $S$ that are 
included in $X$. Dilation of the set $X$ by $S$ is
\begin{equation}
\mathcal{D}_S(X) = \left\{ \mathbf{r} : S_\mathbf{r} \cap  X  \neq 
\emptyset  \right\},
\end{equation}
i.e., all centres of $S$ which hit the set $X$.
Determination of the pore-size distribution is based on morphological opening. 
Opening of the set $X$ by $S$ is erosion followed by dilation, 
\begin{equation}
\mathcal{O}_S(X) = \mathcal{D}_S(\mathcal{E}_S(X)).
\end{equation}
Opening removes small pores from the pore space. Pore-size distribution 
is obtained by successive application of opening by structuring 
element of increasing radius (see, e.g., \citet{Hilpert03}).
Pore volume related to size range $[r,r+\delta r]$ is difference
between the pore volumes determined for pore space opened with structuring 
elements of radius $r$ and $r+\delta r$, i.e., the pore size distribution
is defined as
\begin{equation}
f(r) = \frac{\mathrm{Vol}(\mathcal{O}_{S(r)}(X)) -
  \mathrm{Vol}(\mathcal{O}_{S(r+\delta r)}(X))}{\mathrm{Vol}(X)},
\end{equation}
where $\mathrm{Vol}$ denotes volume of the set in the argument.

Structural anisotropy of biochars was estimated using a method based 
on the grey-scale gradient structure tensor (GST). This method has been used,
e.g., in quantification of anisotropy of trabecular bones from x-ray 
tomography images \citep{Tabor07}. Instead of using the binary image, 
GST was calculated from the denoised grey-scale image. GST analysis requires 
calculation of grey-level gradient at every voxel in the ROI, which 
was obtained by using Sobel's gradient kernels of size  
$5 \times 5 \times 5$ \citep{Pratt07}. The components of GST are obtained 
from the grey-scale gradients as
\begin{equation}
G_{\alpha\beta} = \sum_i g_\alpha(i) g_\beta(i),
\end{equation}
where $g_\alpha(i)$ refers to $\alpha$th component of the grey-level gradient 
at voxel $i$ and summation runs over all image voxels included in the 
ROI. Principal directions are determined as the eigenvectors 
($\mathbf{v}_1$, $\mathbf{v}_2$, $\mathbf{v}_3$) of GST and 
enumerated according to the magnitude of the corresponding eigenvalues 
($\lambda_1 \leq \lambda_2 \leq \lambda_3$). The degree of anisotopy is 
defined as ratio of the highest and lowest eigenvalue,
\begin{equation}
D_A = \frac{\lambda_3}{\lambda_1}.
\end{equation}
High values of $D_A$ indicate high structural anisotropy and values close 
to unity are obtained for isotropic materials. The eigenvector
$\mathbf{v}_1$ indicates the direction of elongated pores.

\subsection{Random walk simulation}

Random walk simulations provide simple method to connect structural 
characteristics of biochars to transport properties. Random walk
was realized as discrete walk in a simple cubic lattice provided by the 
segmented 3D image. The method used here was simular to that described by 
\citet{Nakashima07} and \citet{Promentilla09}.

In a random walk simulation, initial position of the walker is selected
randomly from all pore voxels. At each time step walker moves to one
of the six nearest neighbour voxels. The destination voxel is selected 
randomly, and if the destination is a solid voxel walker stays rest during
that time step. Simulation continues until maximum number of time steps
is exceeded or walker crosses the sample boundary. In this work, the
number of time steps was limited to $10^6$

From each simulation, displacement vector 
$\mathbf{r} = \mathbf{r}_f - \mathbf{r}_0$ was calculated. Here 
$\mathbf{r}_f$ is the position vector of the walker at the end of the 
simulation and $\mathbf{r}_0$ is the initial position. Displacement vector was 
then tranformed to spherical coordinates $(r,\theta,\phi)$ (radial 
distance, azimuthal angle, polar angle).

For each geometry, random walk simulation was repeated $10^5$ times and
the obtained ($\theta$, $\phi$) distribution was compared to the first 
eigenvector of GST in order to connect results of the random walk simulations 
to the structural anisotropy analysis.

\subsection{Biochars}\label{subsec:biochars}

Pyrolysis biochars were produced using batch type bench scale slow pyrolysis 
equipment \citep{Fagernas15}. The reactor was indirectly 
heated with an electric oven and temperature of the oven was controlled via 
a preset program to go through multiple steps. The maximum rate of temperature 
rise was 8$^\circ$C min$^{-1}$. In the final temperature step the oven was 
kept at the selected carbonization temperature for 3 h in order to produce 
fully carbonized biochar. In this study carbonization temperatures of 
375$^\circ$C and 475$^\circ$C were used. 

Hydrothermal carbonization reactions were carried out in a 10 l Hastelloy 
C276 stirred autoclave reactor equipped with an electrical heater band, 
temperature control, digital and analog pressure indicator, internal water 
cooling coil, and PC controlled data logger. The process sequence consisted 
of reactor heat up to 260$^\circ$C (ca. 60 min), residence time of 6 h 
and a water cooling period before venting the residual pressure and opening 
the reactor. 

The studied biochars, processing methods, and temperatures are listed in
Table \ref{table:materials}.
Ash, carbon, oxygen, hydrogen and nitrogen content and pH 
of the biochars used in the study are given in Table \ref{table:properties}.
Loss of ignition at 550$^\circ$C was used to determine the ash content and 
pH was determined at 1:5 biochar to water ratio. Carbon, oxygen, hydrogen 
and nitrogen contents were determined using FLASH 2000 series analyzer. 
In all, biochars included samples of three 
materials (Scots pine bark, salix and coffee cake). Before imaging, biochar 
samples were grinded and sieved to obtain particle size fraction
from 0.84 to 1.19 mm.

\begin{table*}
\caption{List of studied samples including abbreviations, raw material,
processing method and process temperature}
\label{table:materials}
\begin{tabular}{ l l l l }
  \hline\noalign{\smallskip}
  Sample & Material & Process & Temperature [$^\circ$C] \\
  \noalign{\smallskip}\hline\noalign{\smallskip}
  SPB\_P375 & Scots pine bark & Slow pyrolysis & 375 \\
  SPB\_P475 & Scots pine bark & Slow pyrolysis & 475 \\
  SSRC\_P475 & Salix (short rotation coppice) & Slow pyrolysis & 475 \\
  SSRC\_HTC260 & Salix (short rotation coppice) & HTC & 260 \\
  CC\_HTC260 & Coffee cake & HTC & 260 \\
 \noalign{\smallskip}\hline
\end{tabular}
\end{table*}

\begin{table*}
\caption{
Ash (loss of ignition at 550$^\circ$C), carbon, oxygen, hydrogen and 
nitrogen content (\% dry matter) and pH (1:5 biochar to H$_2$O ratio) 
of biochars used in the study. 
}\label{table:properties}
\begin{tabular}{ l l l l l l }
  \hline\noalign{\smallskip}
  Sample & Ash [\%] & C [\%] & O [\%] & H [\%] & pH \\
  \noalign{\smallskip}\hline\noalign{\smallskip}
  SPB\_P375 & 4.4 & 76.4 & 16.3 & 3.8 & 8.0 \\
  SPB\_P475 & 4.6 & 83.7 & 9.3 & 3.0 & 8.4 \\
  SSRC\_P475 & 6.9 & 83.5 & 8.0 & 3.1 & 9.3 \\
  SSRC\_HTC260 & 0.7 & 72.5 & 20.9 & 5.3 & 5.1 \\
  CC\_HTC260 & 0.1 & 74.7 & 15.7 & 7.8 & 4.0 \\
  \noalign{\smallskip}\hline

\end{tabular}
\end{table*}

Since high resolution was used in the imaging, the sample size had 
to be very small due to the tradeoff between the sample size and imaging 
resolution. Sample size was in the sub-millimetre regime and one to three 
biochar grains were imaged of each material. Since materials were 
heterogeneous and their properties varied depending on several other factors, 
e.g., whether the imaged sample grain was from surface or core 
of the processed particle. The raw materials used in biochar production 
were also heterogeneous. For example the Scots pine bark samples 
contained in practice also sapwood xylem in addition to the actual bark. 
It is thus obvious that 
the results reported below do not represent the materials on a general level. 
The purpose here is to find out and highlight the differences and similarities 
of the pore characteristics of several biochars. While the heterogeneity of the
materials complicates interpretation of the results, they represent
realistic biochars used in practical applications.

\section{Results and discussion}\label{sec:results}

\subsection{Visualizations}

\begin{figure*}
  \includegraphics[width=0.95\textwidth]{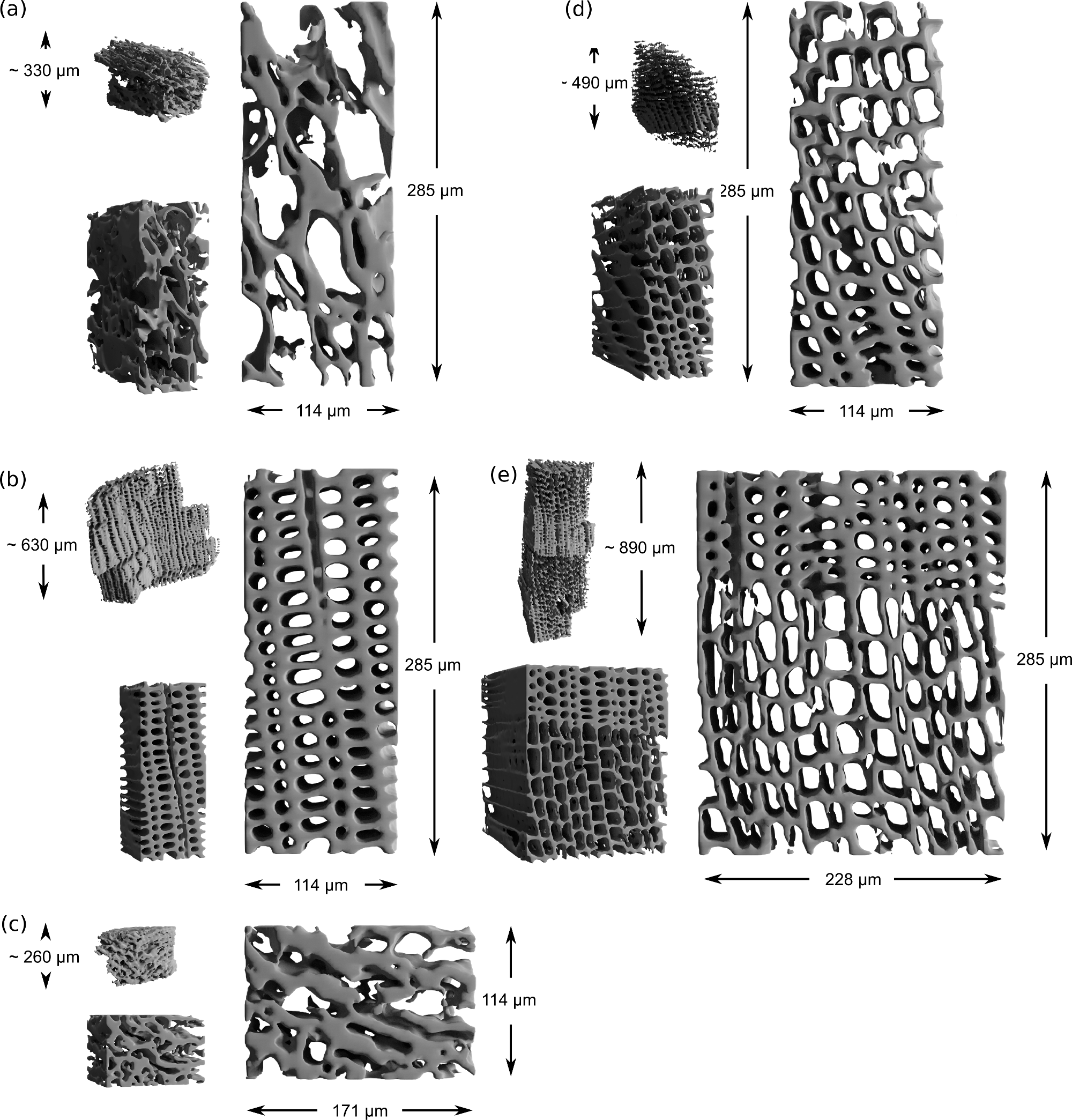}
  \caption{Visualizations of biochars: (a) SPB\_P375\_A, 
(b) SPB\_P375\_B, (c) SPB\_P375\_C, (d) SPB\_P475\_A, (e) SPB\_P475\_B . 
For each case shown is whole ROI at top-left, a smaller cuboidal
piece of the sample at bottom-left and a thin layer taken from the middle of 
the cuboid at right (thickness 57 $\mu$m)
 }\label{fig:vis} 
\end{figure*}

\begin{figure*}
  \includegraphics[width=0.95\textwidth]{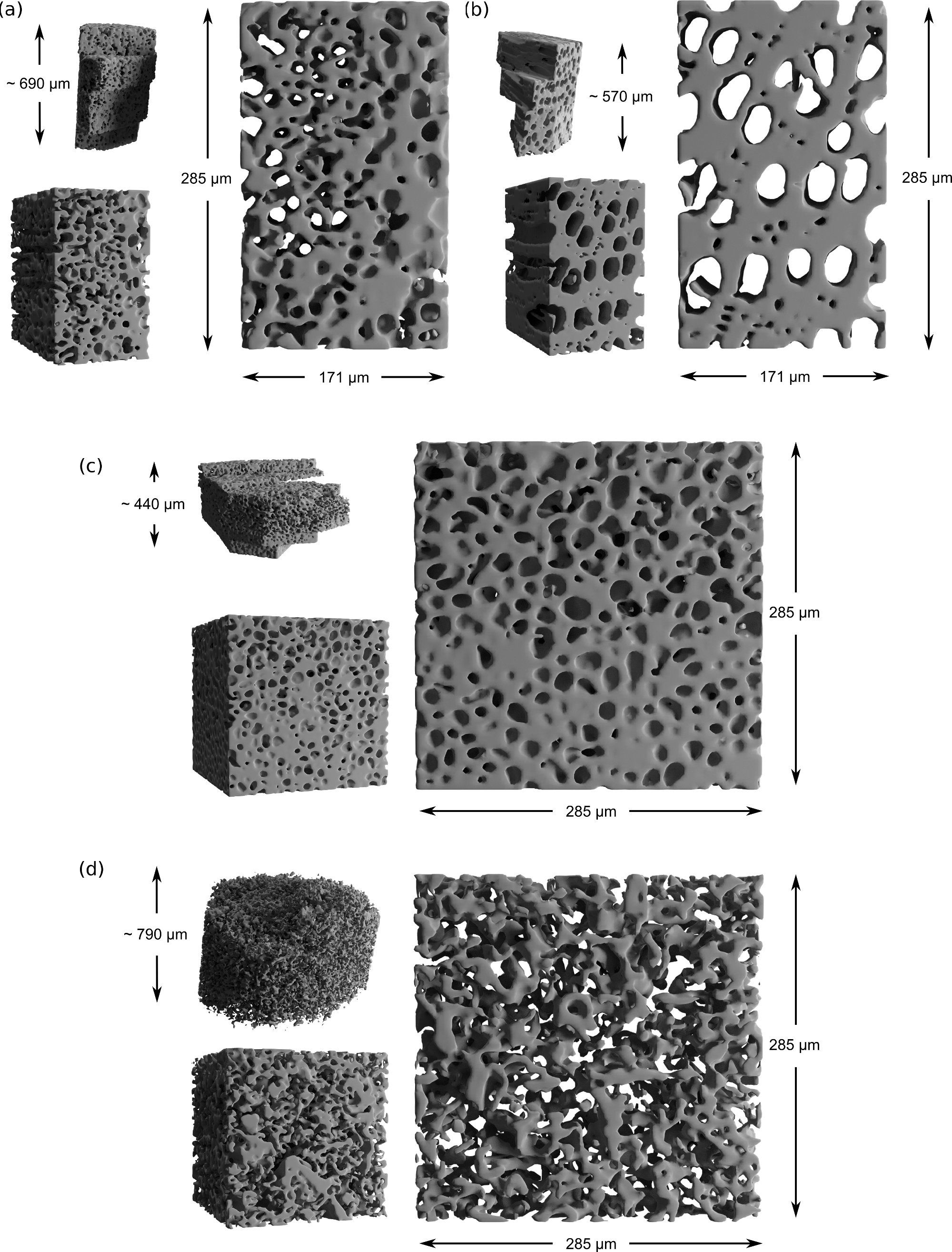}
  \caption{Visualizations of biochars: (a) SSRC\_P475\_A, (b) SSRC\_P475\_B, 
(c) SSRC\_HTC260, (d) CC\_HTC260.
For each case shown is whole ROI at top-left, a smaller cuboidal
piece of the sample at bottom-left and a thin layer taken from the middle of 
the cuboid at right (thickness 57 $\mu$m)
 }\label{fig:vis2} 
\end{figure*}

Visualizations of the imaged samples are shown in Figs. \ref{fig:vis}
and \ref{fig:vis2}. 
For 3D visualizations, the image data was rotated in a way that tube-shaped 
pores were perpendicular to the cutting face. Visual inspection shows that 
there is lot of variation 
in the pore characteristics. The samples can be roughly divided into three
groups. First group contains samples which have highly anisotropic
structure consisting of parallel cylindrical pores with minimal number 
of lateral connections. Such pores result from the xylem structure of 
the wood-based raw material (see Figs. \ref{fig:vis} b,d,e and 
\ref{fig:vis2} b). Samples in the second group have more randomly
shaped and oriented pores (Fig. \ref{fig:vis2} a,c,d). Third group 
consist of two bark samples whose pore structure is an intermediate form
of the two other groups (Fig. \ref{fig:vis} a,c).

The observed differences result from different raw materials used
to produce the biochars. The SSRC material was treated with two different 
technologies, slow pyrolysis and HTC, but the effect of used technology on 
the visual outlook of biochar is not clear. In Fig. \ref{fig:vis2} a and c,
the partially similar visual appearance is obvious, but the pore structure 
of the pyrolysed SSRC sample (Fig. \ref{fig:vis2} b) is different from the 
HTC sample (Fig. \ref{fig:vis2} c). However, it is also important 
to notice that there can be clear differences between different sample 
grains originating from same material. For example, the Scots pine bark 
sample shown in 
Fig. \ref{fig:vis}b is evidently different that the two other bark
samples pyrolysed in $T=375^\circ$C (Fig. \ref{fig:vis} a,c) even though 
the samples were taken from ostensibly same material. The differing sample 
obviously did not represent actual bark as it had clear xylem structure.
Similarly both Scots pine bark samples pyrolysed in $T=475^\circ$C also
originated from sapwood rather than the actual bark layer.
On the other hand, within sample variation due to nature of raw material is 
visible in Fig. \ref{fig:vis} e, where spring and autumn season of annual 
growing ring result in different size of xylem structure. 

On qualitative level there is significant variation in the structural 
properties of different biochars, whereby one might expect that these 
materials behave differently in applications where micron-scale porosity 
plays an important role. These observations indicate that raw material 
homogeneity and quality are important factors to be taken into account 
when biochars are considered for applications where properties of the 
pore structure are essential.

\subsection{Image analyses}

Image analysis results are collected in Table \ref{table:results}. 
Porosities of the samples varied between 0.34 and 0.68, i.e. the highest
porosity was two-fold compared to the lowest one. 
Porosity determination based on imaging obviously takes into account only those
pores that can be observed with the used imaging resolution. Thus the results 
do not contain sub-micron scale porosity which can be essential in some 
applications. 
However, the used resolution was able to account for most of the porosity  
essential for soil hydrological processes and plant growth. The image 
resolution approximately corresponds to pF value 3 if zero contact 
angle is assumed. As biochars are typically 
subcritically hydrophobic \citep{Gray14,Bubici16}, 
the achievable pF value can in practice be somewhat 
lower. Pores in size range down to 100 nm, corresponding to pF values from 
3 to 4.2, are too small to be observed in our tomographic images. Water stored 
in these pores is still considered plant available.

\begin{table*}
\caption{Image analysis results for different samples. Letters A, B and C
refer to distinct grains analysed for each material. MPD and SSA stand for
median pore diameter and specific surface area, respectively
}\label{table:results}
\begin{tabular}{ l l l l l }
  \hline\noalign{\smallskip}
  Sample & Porosity [-] & MPD [$\mu$m] & SSA [mm$^2$/mm$^3$] & $D_A$ [-] \\
  \noalign{\smallskip}\hline\noalign{\smallskip}
  SPB\_P375 & & & & \\
  - A & 0.67 & 22.2 & 96.6 & 6.30 \\
  - B & 0.44 & 12.4 & 124.8 & 33.7 \\
  - C & 0.51 & 12.1 & 126.1 & 8.28 \\
  SPB\_P475  & & & & \\
  - A & 0.55 & 13.8 & 127.6 & 23.4 \\
  - B & 0.60 & 16.2 & 123.5 & 23.1 \\
  SSRC\_P475 & & & & \\
  - A & 0.48 & 13.4 & 132.1 & 1.33 \\
  - B & 0.34 & 19.6 & 87.1 & 21.5 \\
  SSRC\_HTC260 & & & & \\
  - A & 0.42  & 10.7 & 142.1 & 1.76 \\
  CC\_HTC260 & & & & \\
  - A & 0.68 & 13.0 & 126.5 &  1.28 \\
  \noalign{\smallskip}\hline
\end{tabular}
\end{table*}

Due to methodological differences, comparison of the observed porosities to 
those presented in literature is not straightforward. 
\citet{Jones15} detected porosities between 0.23 and 0.27 for cotton hull 
pyrolysed at temperatures of 350, 500 and 800$^\circ$C using tomography with
image resolution of 4 $\mu$m. 
\citet{Gray14} used pycnometry and measured porosities of 83-85 \% and 
63-69 \% for douglas fir and hazelnut shell, respectively, 
pyrolysed at 370, 500 and 620$^\circ$C temperatures. In these studies, 
porosities 
measured for the same raw material pyrolysed in different temperatures showed 
relatively small variation, whereas porosities differed greatly with respect 
to raw material. Similarly, in the present work raw material and 
even within-sample variation seem to be factors determining the porosity. 
In addition, porosity of SSRC material was in the same range irrespectively 
of used processing technology.       

Specific surface area varied from 87 to 142 mm$^2$/mm$^3$. It is interesting 
to observe that highest and lowest specific surface areas were not observed 
for samples with extreme porosity values and that two lowest specific surface 
area values were actually determined for samples with lowest and highest 
porosity. This observation reasserts the impression that there are 
significant differences in the pore-space characteristics of different 
biochars at the length scales visible in tomographic images. 
Specific surface area is strongly affected by the image resolution
which limits the visibility of smaller pores and pore wall roughness.
Thus specific surface area determined from tomographic images
cannot be directly compared to those obtained by gas adsorption
methods. On the other hand, gas adsorption analysis does not provide
information about the pore-size regime essential for using biochars to
improve soil water retention properties.

\begin{figure*}
  \includegraphics[width=0.95\textwidth]{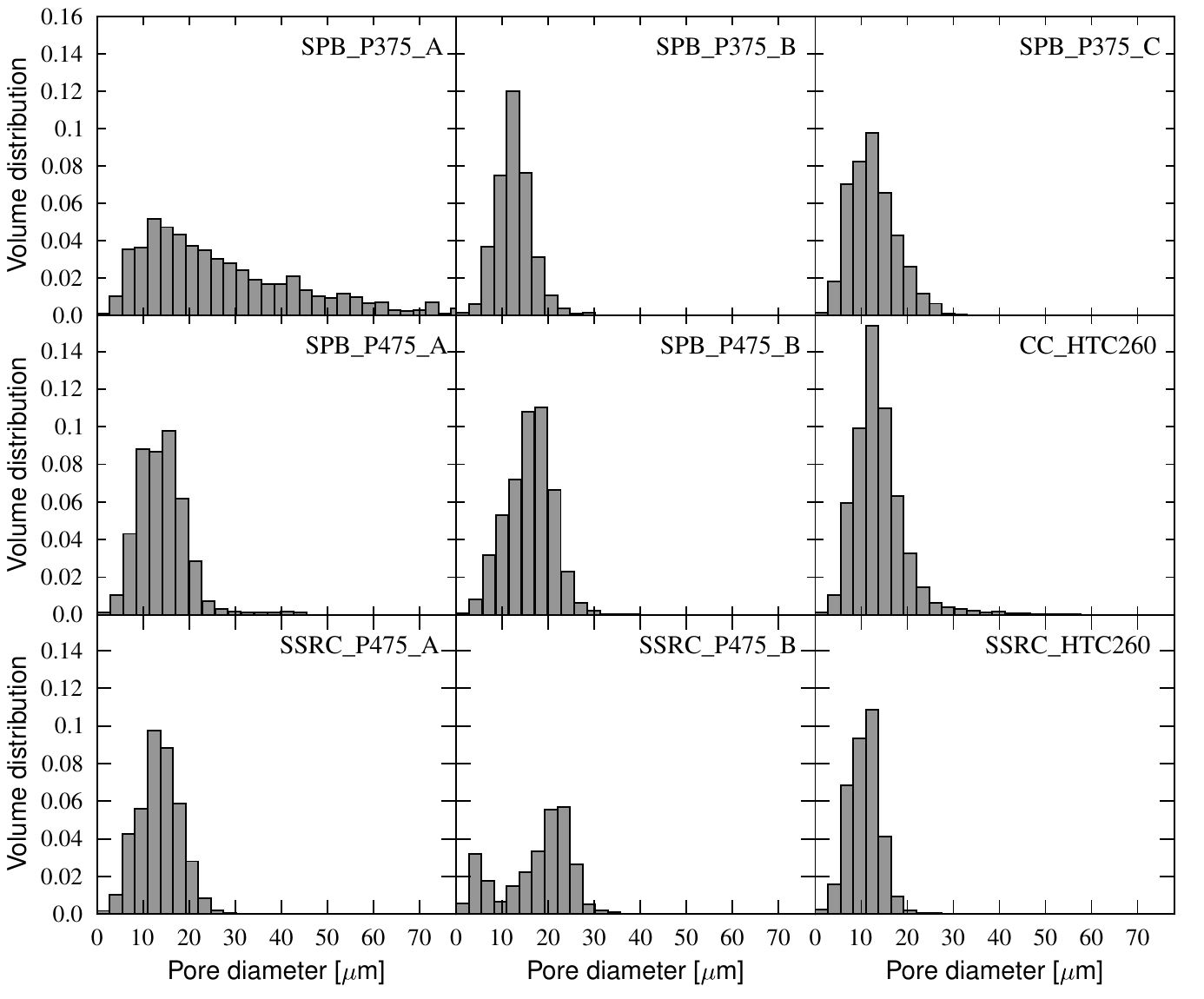}
  \caption{Pore-size distributions. The pore sizes shown on the horizontal
  axes refer to the diameter of the spherical structuring element used in
  the morphological opening}\label{fig:psd} 
\end{figure*}

Pore-size distributions determined for imaged samples are shown in
Fig. \ref{fig:psd}. The pore sizes are quite similar for most of the samples, 
and with one exception distributions have unimodal shape. One pine bark
sample (SPB\_P375\_A) had clearly larger pores, which can be
also observed visually in Fig. \ref{fig:vis} a. The pore-size distribution 
of Scots pine bark sample SPB\_P375\_C is positively skewed and therefore these 
two samples differ from rest of the pine bark samples which do not present
actual bark as already mentioned above.

Salix sample SSRC\_P475\_B
had a bimodal pore size distribution, which is compatible with the visual
observation (Fig. \ref{fig:vis2} b). 
Visual inspection also suggested similarities between the pyrolysed Salix 
sample in Fig. \ref{fig:vis2} a and Salix sample subjected to HTC treatment 
(Fig. \ref{fig:vis2} c). In line with 
that observation, pore size distributions of these two samples are almost 
similar and lack the bimodal structure found in Salix sample SSRC\_P475\_B. 
Once again, raw material quality seems to be key factor determining biochar 
pore system at micrometre scale. This finding also suggests that natural 
variation in the feedstock material sets limits for the extent that biochar 
properties can be optimized by process conditions.   

The dominant pore diameter was in range 10-20 $\mu$m
which corresponds to pF values 2.2-2.5 (assuming zero contact angle).
This pore-size regime contributes to water storage in soils, but is easily 
drained at dry periods allowing gas exchange between larger
inter-particle pores and intra-particle pores of biochar. Thus, this
pore-size regime can contribute actively to many chemical, physical and 
biological phenomena in soils. With respect to agricultural use of biochar, 
amendment with these biochars would increase the capacity of soil to store 
easily plant available water.  

Degree of anisotropy gives a quantitative estimation of the preferential 
orientation of pores. This quantity clearly distinguishes the wood-based 
biochars from the coffee cake sample which had almost isotropic pore structure
($D_A = 1.28$). A clear variation among the Scots pine bark subsamples 
pyrolysed in $T = 375^\circ C$ is observed. Subsamples A and C had clearly 
lower degree of anisotropy than subsample B. The raw material used in
production of biochar contained also wood material that had come off in the 
bark peeling and subsample B clearly represents such material 
(see Fig. \ref{fig:vis} c). Also Scots pine bark sample pyrolysed in 
$T = 475^\circ C$ had high $D_A$ values. Results indicate that the increased
pyrolysis temperature was not able to substantially increase the number of 
micron-scale connections between tracheids. For the wood-based biochars 
eigenvector 
$\mathbf{v}_1$ of the GST was found to be parallel with the orientation of
the xylem (data not shown, see also next section). Samples with xylem
structure had highest $D_A$ values which confirms the visual observations. Thus
$D_A$ can be considered as a good indicator for the anisotropy of the biochar
structure.

No clear differences in the pore characteristics between the Scots pine bark 
samples pyrolysed in $T = 375^\circ C$ (SPB\_P375) and $T = 475^\circ C$ 
(SPB\_P475) could be observed. One could expect some pore development as 
temperature rises (see e.g. \citet{Shaaban14}). However, recent imaging study by
\citet{Jones15} did not show signs of any clear temperature trends. This could
indicate that rising temperature produces pores that are too small to be 
resolved with micrometre resolution. On the other hand, the observed 
heterogeneity of the used biochars would complicate observation of porosity 
development even if such would take place within the studied pore size range.

Either could we see any clear differences in the pore structures of the
salix samples derived by slow pyrolysis (SSRC\_P475) and HTC
(SSRC\_HTC260). Unfortunately also this comparison is
difficult due to the raw material heterogeneity and more significant
differences can be seen between the two pyrolysed subsamples. Some studies
indicate that there are great difference between pyrochars and hydrochars
but this difference is in the regime of small pore sizes not visible
in microtomographic images \citep{Eibisch15}.

\subsection{Random walk simulations}

\begin{figure*}
  \includegraphics[width=0.95\textwidth]{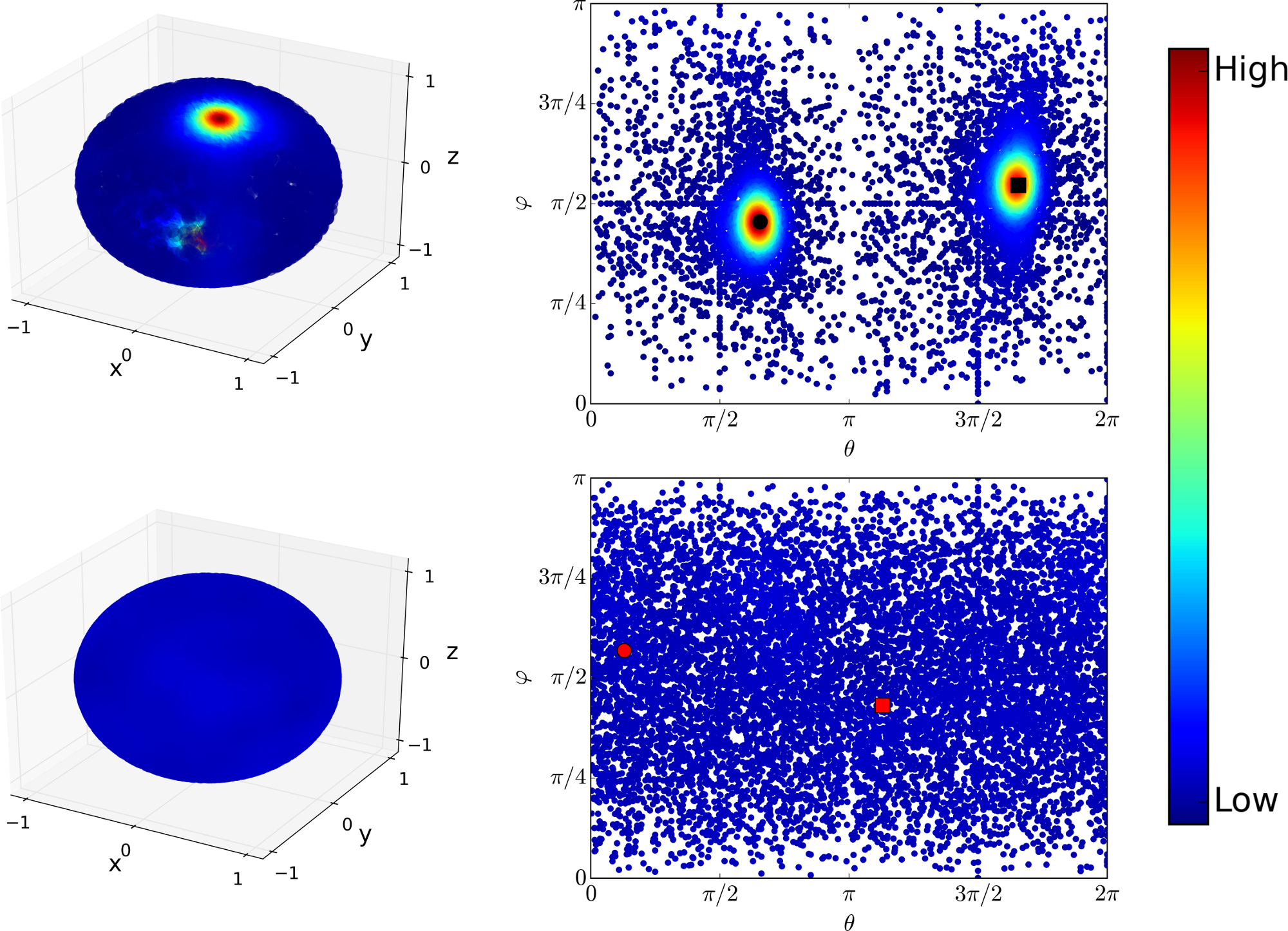}
  \caption{Random walk simulations. Top: SPB\_P375\_B, bottom: CC\_HTC260.
Each point in the figure corresponds to a single random-walk simulation.
On the left side shown are the end points of the unit vectors 
$\mathbf{r}_f -\mathbf{r}_0$.
 On the right side the same points are plotted on the
$\theta\phi$ plane. 
Colour coding shows the spatial density of the points, which is calculated
using the images on the left. Note that same colour coding is used also
for images on the right, i.e. the colour coding is not affected by the
projection distortion. GST results are shown on the right: circle shows the 
direction of the first eigenvector $\mathbf{v}_1$ and square the opposite 
direction ($-\mathbf{v}_1$)
}\label{fig:random}
\end{figure*}

Random walk simulations confirmed the results of structural anisotropy analysis.
Examples of simulation results are shown in Fig. \ref{fig:random}, where
results for SPB\_P375\_B and CC\_HTC260 samples are shown. These two samples
had highest and lowest $D_A$ value.  In the figure,
shown are the orientations of the normalized displacement vectors for all 
$10^5$ simulations performed for both samples. Colour coding shows the spatial
density of the points. Results are notably different for these two materials.
For the highly anisotropic sample (SPB\_P375\_B), displacement vectors 
almost exclusively
follow the direction of eigenvector $\mathbf{v}_1$ (or the opposite direction
-$\mathbf{v}_1$). However, for the most isotropic sample CC\_HTC260, all 
orientations of the displacement vectors are almost equally common as can be 
expected for such an isotropic material. Thus the results of GST analysis
and random walk simulations are very consistent. Random walk simulation
results for other samples were in line with these observations (data not shown).
Simulations confirm that structural anisotropy observed in GST analysis affects
the transport processes (here diffusion) as can be expected.

\section{Conclusions}

The reported results show that x-ray microtomography is a powerful
method to characterize the pore structure of biochars. 3D images enable
quantification of the pore structure by means of image analysis.
Visualizations of the image data provide illustrative information
about the pore structure and therefore helps to gain understanding about
the factors causing the differences in the results of the image analysis.
Thus combination of tomography, visualization techniques and image analyses
provide a functional toolbox for biochar research.

However, feasibility of the methods is limited due to the large amount of work
needed. The use of high image resolution limits the size of the imaged samples 
to sub-millimetre scale, which increases the required number of samples in 
the analyses.
When the studied biochars are made of heterogeneous raw materials, it will
be a real challenge to obtain reliable quantitative information of the pore
characteristic representing whole material. Fortunately also results on single
samples can be highly valuable and interesting even though they do not
necessarily represent whole material. If the methods are used to analyse,
e.g., influence of process methods or conditions on the pore structure,
number of samples could be much smaller if homogeneous sample material is used.

Our results show that considerable fraction of biochar volume consists 
of pores in size classes relevant for storage of plant available water. Thus 
direct effects of biochars on soil water retention cannot be ruled out.
Also the prevailing custom to characterize biochar pore space only with 
gas adsorption analyses is clearly insufficient, especially if biochar 
use aims to improve water retention and storage, or gas exchange in soils.

\begin{acknowledgements}
This project has received funding from the European Union's Horizon 2020 
research and innovation programme under grant agreement 
No 637020 -- MOBILE FLIP.
\end{acknowledgements}

\end{document}